\documentclass[final,authoryear,5p,times,twocolumn]{elsarticle}

\usepackage{graphicx}

\journal{New Astronomy}

\usepackage{amssymb}

\def\astrobj#1{#1}

\begin{document}

\begin{frontmatter}

\title{Extensive photometry of the WZ~Sge-type dwarf nova \astrobj{V455~And} (\astrobj{HS2331+3905}): detection of negative superhumps and coherence  features of the short-period oscillations}

\author{V. P. Kozhevnikov\corref{cor1}},

\ead{valery.kozhevnikov@urfu.ru}

\cortext[cor1]{Tel.: + 7 343 2615431; fax: + 7 343 3507401.}

\address{Astronomical Observatory, Ural Federal University, Lenin Av. 51,
Ekaterinburg, 620083, Russia}

\begin{abstract}

We report the results of photometry of the WZ~Sge-type dwarf nova \astrobj{V455 And}. Observations were obtained over 19 nights in 2013 and 2014. The total duration of the observations was 96~h. We clearly detected three coherent oscillations with periods of $80.376\pm0.003$~min, $40.5431\pm0.0004$~min and $67.619685\pm0.000040$~s. The 67.619685-s period can be the spin period of the white dwarf. The 40.5431-minute period is the first harmonic of the orbital period. The 80.376-minute oscillation can be a negative superhump because its period is 0.9\% less than the orbital period. This oscillation was evident both in the data of 2013 and in the data of 2014. These results make \astrobj{V455 And} a permanent superhump system which shows negative superhumps. This is also the first detection of persistent negative superhumps in a WZ Sge-type dwarf nova. In addition, the analysis of our data revealed incoherent oscillations with periods in the range 5--6 min, which were observed earlier and accounted for by non-radial pulsations of the white dwarf. Moreover, we clearly detected an oscillation with a period of $67.28\pm0.03$~s, which was of a low degree of coherence. This oscillation conforms to the beat between the spin period of the white dwarf and the 3.5-h spectroscopic period, which was discovered earlier and accounted for by the free precession of the white dwarf. Because the 67.28-s period is shorter than the spin period and because the free precession of the white dwarf is retrograde, we account for the 67.28-s oscillation by the free precession of the white dwarf.

\end{abstract}


\end{frontmatter}

\section{Introduction}

Cataclysmic variables (CVs) are interacting binaries, in which a white dwarf accretes material from a late-type companion filling its Roche lobe. Many CVs of short orbital period have light curves with prominent humps, which reveal periods slightly longer than $P_{\rm orb}$. These are called superhumps, as they are characteristic of SU UMa-type dwarf novae in superoutbursts. Some bright nova-like variables and nova remnants show superhumps during the normal brightness state. These are called permanent superhumps. Their periods can be either a few per cent longer than $P_{\rm orb}$, in which case they are called positive superhumps, or a few per cent shorter, in which case they are called negative superhumps \citep[e.g.,][]{retter00,retter02}.

Intermediate polars (IPs) form a subclass of CVs, in which a magnetic white dwarf accretes material from a late-type companion through a truncated accretion disc. The rotation of the white dwarf is not phase-locked to the binary period of the system. Because the magnetic axis is offset from the spin axis of the white dwarf, this causes oscillations in the X-ray and optical wavelength bands. The X-ray oscillation period is usually identified as the spin period of the white dwarf. In addition to the spin and orbital periods, the reprocessing of X-rays in some part of the system that rotates with the orbital period gives rise to emission that varies with the beat period, where $1/P_{\rm beat}=1/P_{\rm spin}-1/P_{\rm orb}$. This synodic counterpart is often called the orbital sideband of the spin frequency. A comprehensive review of IPs is given in Patterson (1994).

The cataclysmic variable \astrobj{HS 2331+3905} was discovered in the Hamburg Quasar Survey \citep{hagen95}. Lately it was renamed \astrobj{V455 And}. Follow-up observations were accomplished by \citet{araujo05}. They defined \astrobj{V455 And} as a CV with a very wide range of variability.  \citeauthor{araujo05} reported the detection of an orbital period of 81.08 min, a superhump period of 83.38 min and non-radial pulsations of the white dwarf with periods in the range 5--6 min. In addition, they detected the coherent signal with a period of 1.12 min. This signal may be caused by the spin of the magnetic white dwarf. The spectroscopic period found by them (3.5~h) turned out to be inconsistent with the orbital period defined by the eclipses. This spectroscopic period was not strictly coherent and drifted in a time-scale of days. Moreover, \citeauthor{araujo05} showed that the orbital light curve of \astrobj{V455~And} was similar to the orbital light curve of the well-studied dwarf nova \astrobj{WZ Sge}. WZ Sge-type stars exhibit extremely bright superoutbursts every 2--3 decades. Such a superoutburst in \astrobj{V455~And} happened in 2007 September. Thus, \astrobj{V455 And} was classified as a WZ Sge-type system \citep{ritter03}.

A few features of \astrobj{V455 And} seem quite puzzling. The first puzzle concerns the superhumps with a period of 83.38 min. Because \citeauthor{araujo05} detected the superhumps during the quiescence of \astrobj{V455 And}, these superhumps must be treated as permanent positive superhumps. However, it is known that positive superhumps appear in short-period CVs that reach high brightness states \citep{patterson05}. The WZ Sge stars in quiescence, however, have extremely low brightness. The second puzzle concerns the 1.12-minute oscillation, which may be connected with the spin of the white dwarf. \cite{gansicke07} discovered that this oscillation consists of two components with periods of 67.62 s and 67.24 s. These periods are not connected with each other by the relation, $1/P_{\rm beat}=1/P_{\rm spin}-1/P_{\rm orb}$, which is valid in classical IPs. The 67.24-s period appears as the beat between the 67.62-s period and 3.5-h spectroscopic period. This spectroscopic period is in no way related to the orbital period and also seems quite puzzling. Moreover, \citeauthor{gansicke07} discovered that the 67.24-s oscillation is incoherent. Until recently no conceivable reason was proposed for this oscillation.

To investigate the properties of the short-period oscillations that were observed in \astrobj{V455 And}, we performed extensive photometric observations and found that the high-frequency component of the 1.12-minute oscillation is of a low degree of coherence. We suggest a conceivable reason for this oscillation. Moreover, instead of the positive superhumps detected by \citeauthor{araujo05}, we clearly detected negative superhumps. In this paper, we present the results of all our observations, spanning a total duration of 96~h within 19 nights.

\section{Observations}

\begin{table}
\caption[ ]{Journal of the observations}
\label{journal}
\begin{flushleft}
\begin{tabular}{lll}
\noalign{\smallskip}
\hline
\noalign{\smallskip}
{\hspace{9mm}}Date & {\hspace{7mm}} HJD start & {\hspace{7mm}} Length \\
{\hspace{8mm}}(UT) & {\hspace{6mm}} (-2,456,000) & {\hspace{6mm}} (h) \\
\noalign{\smallskip}
\hline
\noalign{\smallskip}
2013 September 12 & {\hspace{8mm}}  548.254941 & {\hspace{8mm}}  5.4  \\
2013 September 13 & {\hspace{8mm}} 549.226710  & {\hspace{8mm}}  2.6 \\
2013 October  8      & {\hspace{8mm}}  574.141304 & {\hspace{8mm}}  2.2 \\
2013 October 10     & {\hspace{8mm}}  576.133304 & {\hspace{8mm}}  5.5 \\
2013 November 7   & {\hspace{8mm}}  604.101156 & {\hspace{8mm}}  6.7 \\
2013 November 8   & {\hspace{8mm}}  605.243913 & {\hspace{8mm}}  7.1 \\
2013 November 21 & {\hspace{8mm}}  618.074679 & {\hspace{8mm}}  2.3 \\
2013 November 23 & {\hspace{8mm}}  620.068194 & {\hspace{8mm}}  4.9 \\
2013 November 24 & {\hspace{8mm}}  621.066324 & {\hspace{8mm}}  1.5 \\
2013 November 25 & {\hspace{8mm}}  622.098547 & {\hspace{8mm}}  3.3 \\
2013 November 26 & {\hspace{8mm}}  623.063014 & {\hspace{8mm}}  5.9 \\
2013 December 4   & {\hspace{8mm}}  631.082257 & {\hspace{8mm}}  4.6 \\
2014 January 3        & {\hspace{8mm}}  661.074162 & {\hspace{8mm}}  5.7 \\
2014 January 4        & {\hspace{8mm}}  662.095586 & {\hspace{8mm}}  5.9 \\
2014 September 22 & {\hspace{8mm}}  923.195647 & {\hspace{8mm}}  7.0 \\
2014 September 23 & {\hspace{8mm}}  924.194650 & {\hspace{8mm}}  3.0 \\
2014 September 24 & {\hspace{8mm}}  925.166716 & {\hspace{8mm}}  8.1 \\
2014 September 25 & {\hspace{8mm}}  926.172375 & {\hspace{8mm}}  7.2 \\
2014 September 27 & {\hspace{8mm}}  928.160956 & {\hspace{8mm}}  7.2 \\

\noalign{\smallskip}
\hline
\end{tabular}
\end{flushleft}
\end{table}

In observations of CVs, we use a multi-channel photometer with photomultiplier tubes that allows us to make continuous brightness measurements of two stars and the sky background. Because the angular separation between the programme and comparison stars is small, such differential photometry allows us to obtain magnitudes, which are corrected for first-order atmospheric extinction and for other unfavourable atmospheric effects (unstable atmospheric transparency, light absorption by thin clouds, etc.). Moreover, we use the CCD guiding system, which enables precise centring of the two stars in the diaphragms to be maintained automatically. This greatly facilitates the acquisition of long continuous light curves and improves the accuracy of brightness measurements. The design of the photometer is described in \citet{kozhevnikoviz}.

The main part of our photometric observations of \astrobj{V455 And} was obtained in 2013 September--December and in 2014 January over 14 nights using the 70-cm telescope at Kourovka Observatory, Ural Federal University. Below, for brevity, the data obtained in 2014 January are analysed together with the data obtained in September through December 2013. In 2014 September over 5 nights we obtained additional photometric observations of \astrobj{V455 And} with the goal to support spectroscopic observations of \astrobj{V455 And} at Terskol Observatory, Russia. Because these additional data are separated by too large a gap from the main part of the data, this results in uncertainty in the window function. Therefore, we did not include these additional data into the common time series. Instead, we used them for confirmation of the results obtained from the data of 2013. A journal of the observations is given in Table~\ref{journal}.

\begin{figure}[t]
\resizebox{\hsize}{!}{\includegraphics{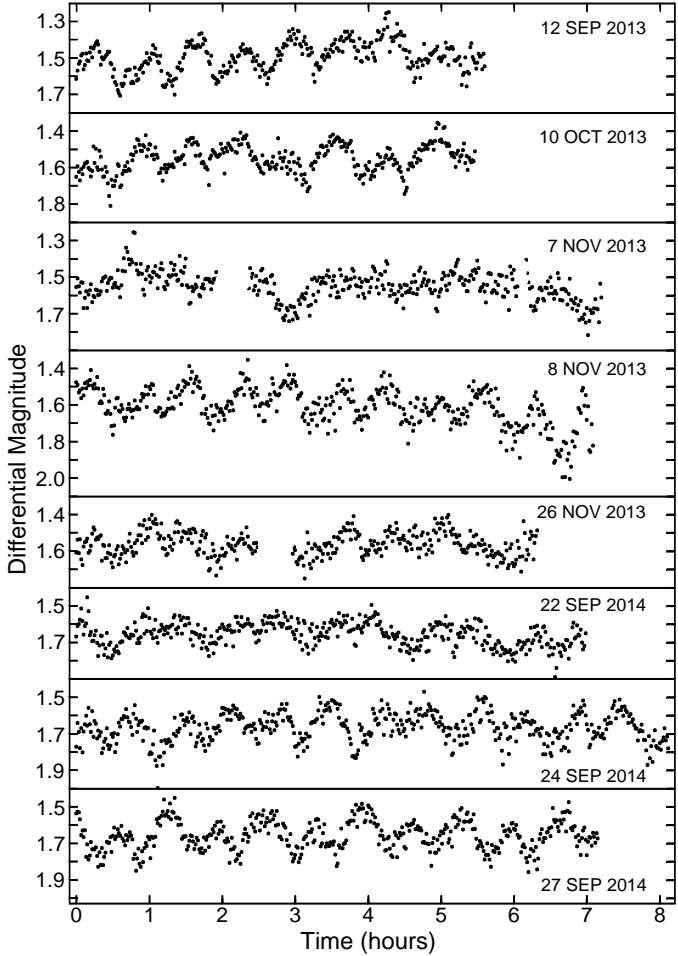}}
\caption{Longest differential light curves of \astrobj{V455 And}.}
\label{figure1}
\end{figure}

The programme and comparison stars were observed through 16-arcsec diaphragms, and the sky background was observed through a 30-arcsec diaphragm. The comparison star is USNO-A2.0 1275-18482440. It has $\alpha=23^h33^m40.79^s$, $\delta=+39^\circ17'24.6''$ and $B=14.7$~mag. Data were collected at 1-s sampling intervals in white light (approximately 3000--8000~\AA), employing a PC-based data-acquisition system. We obtained differences of magnitudes of the programme and comparison stars taking into account the differences in light sensitivity between the various channels. According to the mean counts, the photon noise (rms) of the differential light curves is equal to 0.17--0.26~mag (a time resolution of 1~s). The actual rms noise also includes atmospheric scintillations and the motion of the star images in the diaphragms. But these noise components give only insignificant additions. Fig.~\ref{figure1} presents the longest differential light curves of \astrobj{V455 And}, with magnitudes averaged over 60-s time intervals. The white-noise component of these light curves is 0.022--0.034~mag.

\section{Analysis and results}

Observations with multichannel photometers provide data evenly spaced in time. Then, in cases of smooth signals, classical methods of analysis such as the Fourier transform turn out optimal in comparison with numerous methods appropriate to unevenly spaced data \citep[e.g., the Lomb-Scargle periodogram,][]{schwarzenberg98}. In addition, the Fourier transform is preferential in comparison with methods based on phase analysis of folded light curves, \citep[e.g., the analysis-of-variance method (AOV),][]{schwarzenberg89} because, in cases of multiperiodicity, different periods can be well separated in a Fourier power spectrum, whereas this can be difficult in an AOV periodogram due to subharmonics produced at multiple periods.

\begin{figure*}[t]
\resizebox{\hsize}{!}{\includegraphics{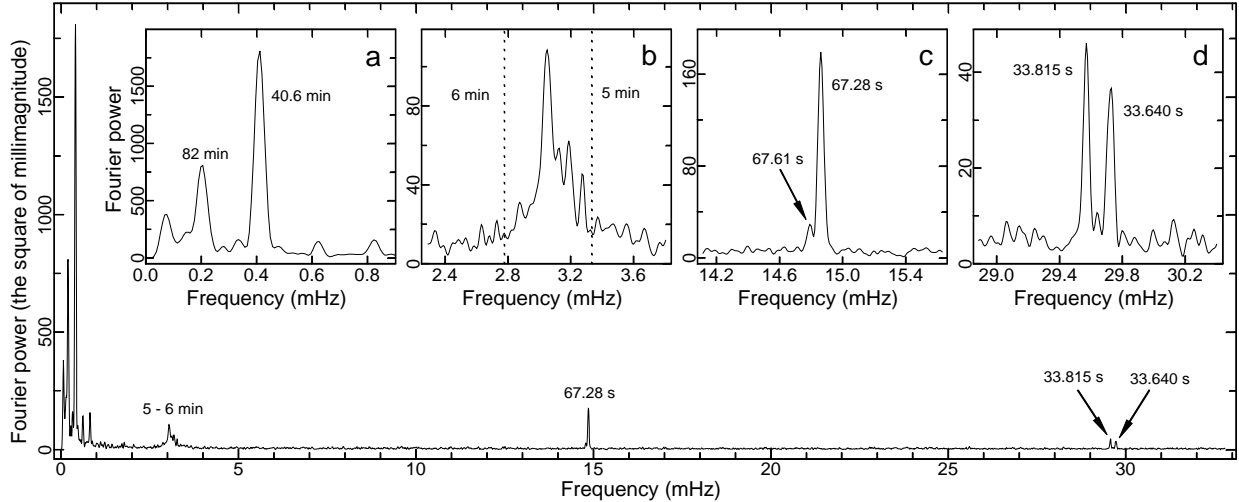}}
\caption{ Average power spectrum calculated by the averaging of 9 power spectra of longest individual light curves of 2013 from \astrobj{V455 And}. It shows various periodic oscillations. The inserts show the corresponding parts of the power spectrum on a large scale.}
\label{figure2}
\end{figure*}

A Fourier power spectrum of a continuous time series allows us to detect a coherent oscillation due to a sharp peak. Then, the full width of the peak at half-maximum (FWHM) is roughly equal to the frequency resolution $1/T$, where $T$ is the length of the time series.  At first, we calculated individual power spectra, which allowed us to detect signals, which are coherent during a night. Fig.~\ref{figure2} presents the average power spectrum obtained by the averaging of 9 power spectra calculated by a fast Fourier transform algorithm for the longest light curves obtained in 2013. Previously, low-frequency trends were removed from the light curves by subtraction of a second-order polynomial fit. This power spectrum reveals sharp peaks with periods of $82\pm6$ and $40.6\pm0.6$~min (the insert~a), 6$7.61\pm0.06$ and $67.28\pm0.03$~s (the insert~c), $33.815\pm0.012$ and $33.640\pm0.017$~s (the insert~d). Two latter oscillations obviously correspond to the first harmonics of the oscillations, the peaks of which are seen in the insert~c.  The widths of the peaks in the inserts a, c and~d conform to the coherence of the corresponding oscillations.

Here, the errors are equal to the half-widths of the peaks in the power spectrum at the $p-N^2$ level, where $p$ is the peak height and $N^2$ is the mean noise power level. These errors are found according to \citet{schwarzenberg91}. We, however, applied the method by \citeauthor{schwarzenberg91} to an average power spectrum, the statistical properties of which are better than statistical properties of a single power spectrum. Therefore, these errors are conservative errors but not rms errors.

\begin{figure}[hptb]
\resizebox{\hsize}{!}{\includegraphics{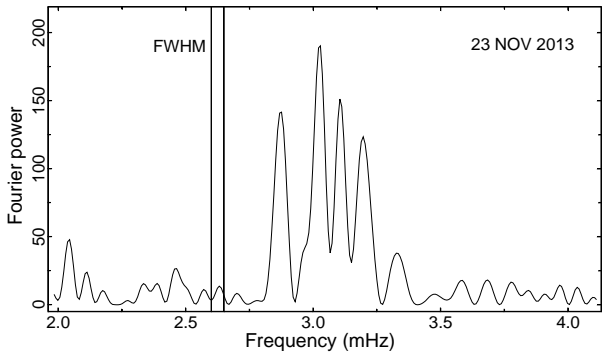}}
\caption{Part of the power spectrum calculated for the light curve of 2013 November 23 from \astrobj{V455 And}. Two vertical solid lines mark FWHM of the peak of an oscillation coherent within the entire light curve (4.9~h). This FWHM is obtained from an artificial time series consisting of a sine wave.}
\label{figure3}
\end{figure}

In contrast to the inserts a, c and d, the insert b in Fig.~\ref{figure2} reveals a much wider peak between 5 and 6 min. This means either an oscillation of a low degree of coherence or a cumulative effect of several coherent oscillations. A similar picture was also found in the average power spectrum obtained by the averaging of 5 power spectra of the light curves of 2014. Carefully examining the individual power spectra, we made sure that the latter case is roughly true. Fig.~\ref{figure3}  shows the part of the  power spectrum of one of the individual light curves of \astrobj{V455 And}, in which four peaks with periods between 5 and 6 min are seen. Only one of them reveals FWHM corresponding to FWHM of an oscillation coherent within the entire light curve (4.9~h). Other peaks reveal wider FWHM. Thus, we conclude that only rare oscillations with periods between 5 and 6 min can be coherent within 5 hours. The coherence time of most other oscillations can be less than 5 hours. Moreover, a few oscillations with periods between 5 and 6 min can coexist simultaneously.

To establish the coherence of oscillations during large time intervals, one can analyse data incorporated into common time series. Then the coherence can be demonstrated due to coincidence of the structure of the power spectrum in the vicinity of the oscillation frequency and the window function. In the simplest case, the window function is characterized by the presence of 1-d aliases. When a common time series contains large gaps, the window function reveals the fine structure, which must be evident in the power spectrum. Obviously, the highest precision of oscillation periods can be achieved from all data incorporated into common time series. However, too large gaps can result in uncertainty of the window function and aliasing problem.

\begin{figure*}[t]
\resizebox{\hsize}{!}{\includegraphics{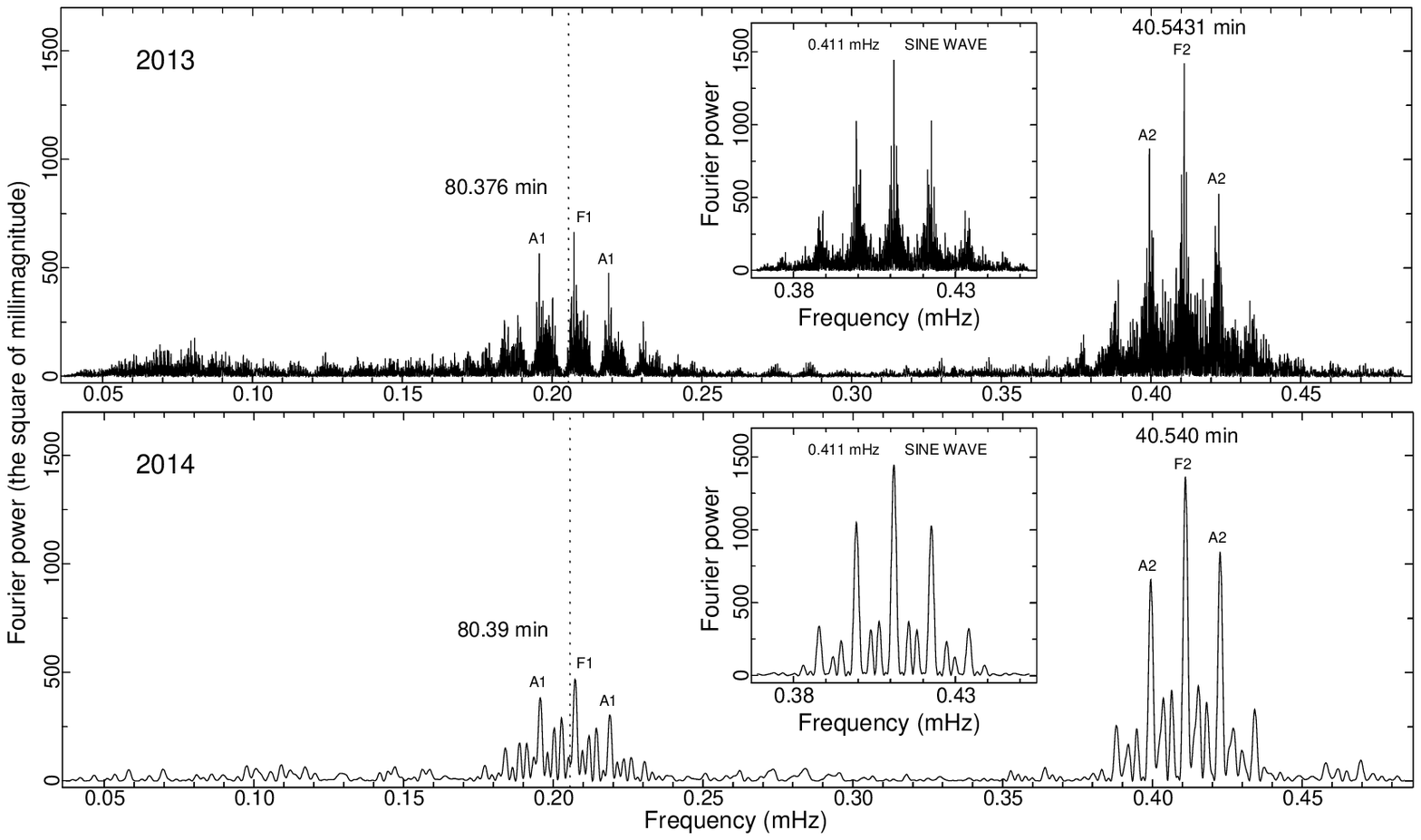}}
\caption{Low-frequency parts of the power spectra calculated for two groups of data from \astrobj{V455~And}. The Inserts show the window functions. The principal peaks are labelled with ''F1'' and ''F2'', and the 1-d aliases are labelled with ''A1'' and ''A2''. The dotted line marks the orbital period.}
\label{figure4}
\end{figure*}

Fig.~\ref{figure4} presents the Fourier power spectra of two common time series consisting of the data of 2013 and 2014 in the vicinity of the oscillations with periods of 82 and 40.6 min. The gaps due to daylight and poor weather in these time series were filled with zeros. Previously, low-frequency trends were removed from the individual light curves by subtraction of a first- or second-order polynomial fit. As seen, the structures of the power spectra are remarkably similar to the window functions shown in the inserts. In addition to 1-d aliases, the window functions reveal fine structures. Note that the window function corresponding to the data of 2013 is characterized by the presence of a few additional peaks disposed very close to the principal peak and 1-d aliases. Coincidence of their locations and the locations of the additional peaks in the power spectrum of the real data is especially noteworthy. Obviously, both oscillations are coherent both within the data of 2013 and within the data of 2014. This proves the reality of these oscillations.

The precise maxima of the principal peaks in the power spectra presented in Fig.~\ref{figure4} were found by a Gaussian function fitted to the upper parts of the peaks. We obtained the most precise periods of two oscillations from the data of 2013, namely $P_{\rm 1}=80.376\pm0.003$ min and $P_{\rm 2}=40.5431\pm0.0004$ min. The data of 2014 gave the less precise periods, namely $P_{\rm 1}=80.39\pm0.07$ min and $P_{\rm 2}=40.540\pm0.010$ min. In this case, the lesser precision is caused by the lesser observational coverage. The errors are found according to \citet{schwarzenberg91}. Because we used single power spectra, these errors are rms errors or $\sigma$. Indeed, the difference of the periods obtained from the data of 2013 and  the data of 2014 is in the range 0.2--0.3$\sigma$. This is consistent with the rule of 3$\sigma$. In addition, this shows that these two oscillations have stable periods.

\begin{figure}[hptb]
\resizebox{\hsize}{!}{\includegraphics{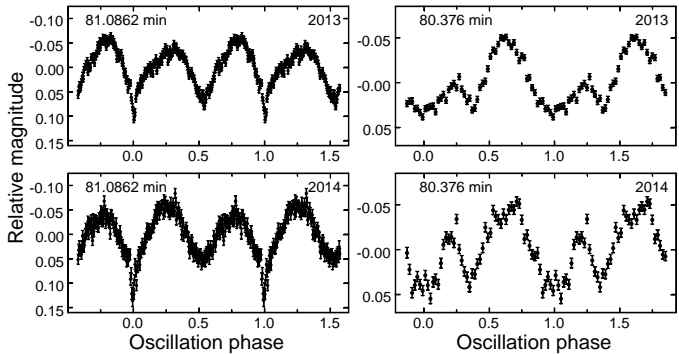}}
\caption{Light curves of \astrobj{V455 And} folded with a period of 81.0862 min (on the left) and with a period of 80.376 min (on the right).}
\label{figure5}
\end{figure}

As discovered by \citet{araujo05}, the oscillation with $P_{\rm 2}$ is the first harmonic of the orbital period of \astrobj{V455 And}. Therefore, to find the pulse profile of orbital variability, we folded the light curves with the doubled period. Indeed, as seen in Fig.~\ref{figure5} on the left, even and odd cycles of this folded light curve differ from each other by a shallow eclipse. This eclipse demonstrates that $P_{\rm 2}$ is the first harmonic of $P_{\rm orb}$. Thus, we obtained $P_{\rm orb}=81.0862\pm0.0008$ min. As seen in Fig.~\ref{figure5}, the profile of the orbital light curve is only slightly changeable during the year of our observations.

The period $P_{\rm 1}$ obtained from the data of 2013 differs from  $P_{\rm orb}$. The difference is $0.710\pm0.003$ min or $240\sigma$ and therefore is undoubted. The data of 2014 gives the difference of these periods, which is equal to $0.69\pm0.07$ min or $10\sigma$. The differences of these two periods are compatible with each other. The period $P_{\rm 1}$ is 0.9\% less than $P_{\rm orb}$, and, therefore, $P_{\rm 1}$ can be treated as a negative superhump period. In Fig.~\ref{figure4}, we marked $P_{\rm orb}$ by the dotted line. This visually demonstrates that $P_{\rm 1}$ is less than $P_{\rm orb}$ both in 2013 and in 2014. Thus, the superhump period, $P_{\rm sh}$, is equal to $80.376\pm0.003$ min.

The light curves folded with $P_{\rm sh}$ are presented in Fig.~\ref{figure5} on the right. Both the light curve obtained from the data of 2013 and  the light curve obtained from the data of 2014 reveal similar double-humped pulse profiles with small interhumps between main humps. Thus, not only the superhump period but also the superhump pulse profile remained practically unchangeable during the year of our observations.

The power spectra (Fig.~\ref{figure4}) show enlarged noise levels in the vicinity of $P_{\rm sh}$  and in the vicinity of the first harmonic of $P_{\rm orb}$. This suggests the presence of additional oscillations. To find this out, we used the well-known method of subtraction of the largest oscillation from data. In the case of $P_{\rm sh}$, the residual power spectrum revealed chaotically situated noise peaks with heights less than heights of noise peaks at the lowest frequencies. Hence, an additional detectable coherent oscillation is absent in the vicinity of $P_{\rm sh}$. Moreover, we conclude that the orbital variability of \astrobj{V455 And} produces no fundamental oscillation. In contrast, in the case of the first harmonic of $P_{\rm orb}$, the residual power spectrum revealed a coherent oscillation with a period of $40.1871\pm0.0011$~min. Of course, it is the first harmonic of $P_{\rm sh}$, which arises due to the nonsinusoidal pulse profile of the superhumps (Fig.~\ref{figure5}). This first harmonic additionally confirms the reality of the negative superhumps.

We checked the influence of pre-whitening upon the folded light curves presented in Fig.~\ref{figure5} and found no considerable effects. Obviously, the folded light curves are obtained by using sufficient portions of data, and undesirable effects, which might be caused by other oscillations having close frequencies, are supressed.

The data incorporated into common time series did not allow us to detect coherent oscillations in the range 5--6 min. The power spectrum of the common time series consisting of the data of 2013 and the gaps according to the observations repeated the contour of the wide peak seen in the insert b of Fig.~\ref{figure2} and showed no signs of the window function. This means that either the oscillations are coherent but too numerous or they are incoherent from night to night. The latter case seems most probable because it conforms to the low coherence time of the oscillations detected in the individual light curve (Fig.~\ref{figure3}).

\begin{figure}[hptb]
\resizebox{\hsize}{!}{\includegraphics{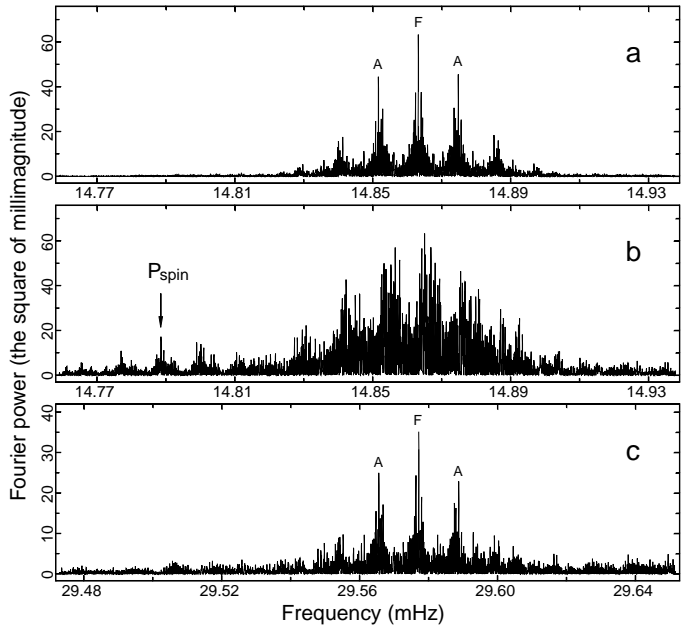}}
\caption{ Parts of the power spectrum calculated for the data of 2013 from \astrobj{V455 And}. They are shown in the vicinity of the 67.28-s oscillation (b) and in the vicinity of the first harmonic of the 67.61-s oscillation (c). The window function is shown in the frame a. The principal peaks are labelled with ''F'', and the 1-d aliases are labelled with ''A''.}
\label{figure6}
\end{figure}

\begin{figure}[t]
\resizebox{\hsize}{!}{\includegraphics{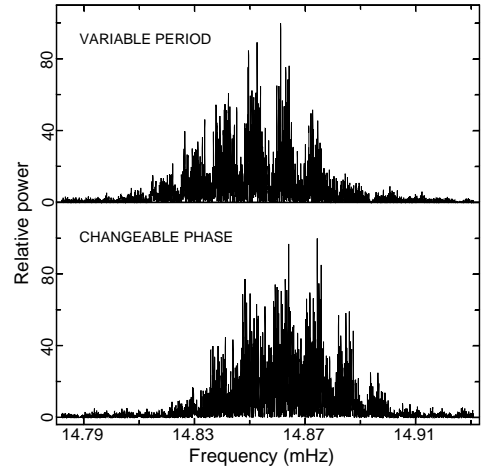}}
\caption{Power spectra of two artificial time series, which simulate the oscillation with a period of 67.28 s according to the observations in 2013. The upper frame shows the effect of a period varying with ${\rm d}P/{\rm d}T=10^{-8}$.  The lower frame shows the effect of an unstable phase, which fluctuates with a standard deviation of $20\%$ of the period from night to night.}
\label{figure7}
\end{figure}

The oscillations with periods of 67.61 and 67.28~s seem very interesting because one of the periods can be the spin period of the white dwarf, whereas another period can be a beat period. In classical IPs, spin oscillations are extremely coherent. However, our power spectrum of the common time series consisting of the data of 2013 and the gaps according to the observations (Fig.~\ref{figure6}) showed a very unusual picture.  Indeed, the oscillation with a period of 67.61~s is coherent. This is especially clearly seen in its first harmonic (Fig.~\ref{figure6}c). In contrast, the oscillation with a period of 67.28~s is of a low degree of coherence. As seen in Fig.~\ref{figure6}b, many thin peaks are clustered around the places spaced as the principal peak and 1-d aliases. This does not conform to the window function presented in Fig.~\ref{figure6}a. None the less, this may mean that this oscillation is coherent within a few close nights. 

\begin{figure}[t]
\resizebox{\hsize}{!}{\includegraphics{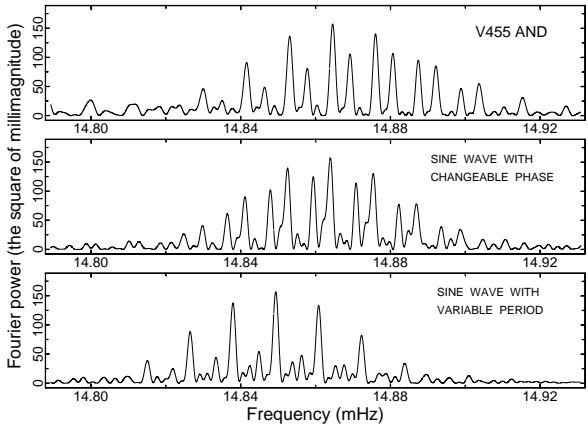}}
\caption{Power spectrum calculated for the data of 2013 November 21--26 from \astrobj{V455 And} (the upper frame) and power spectra of two artificial time series, which simulate the oscillation with a period of 67.28 s according to the observations in 2013 November 21--26. The middle frame shows the effect of an unstable phase, which fluctuates with a standard deviation of 20\% of the period from night to night. The lower frame shows the effect of a period varying with ${\rm d}P/{\rm d}T=10^{-8}$.}
\label{figure8}
\end{figure}

Because the 67.61-s period is coherent within a sufficiently long time span, we can consider it the spin period of the white dwarf. From the power spectrum shown in Fig.~\ref{figure6}b, we found $P_{\rm spin}=67.619685\pm0.000038$~s. The first harmonic of this oscillation allows us to achieve the more precise value of $P_{\rm spin}$. The period of the first harmonic is equal to $33.8098585\pm0.0000070$~s. Thus, finally, we found $P_{\rm spin}=67.619717\pm0.000014$~s.

To find possible reasons for the low degree of coherence of the 67.28-s oscillation, we performed numerical experiments which simulated this oscillation. We constructed two artificial time series consisting of a sine wave with a period of 67.28~s and with the gaps according to the observations in 2013. In the first case, the period of the sine wave varied with ${\rm d}P/{\rm d}T=10^{-8}$. In the second case, the phase of the sine wave fluctuated with a standard deviation of 20\% of the period from night to night. As seen in Fig.~\ref{figure7}, the power spectra of these time series reveal the pictures similar to the pictures really observed for this oscillation, i.e. many thin peaks are clustered around the places spaced as the principal peak and 1-d aliases. 

To realise which of the two cases occurs, we confined the length of the artificial time series according to five nearly consecutive nights in 2013 November 21--26. The middle frame of Fig.~\ref{figure8} shows the power spectrum of the artificial time series, in which the  phase of the sine wave fluctuated with a standard deviation of 20\% of the period from night to night. This power spectrum imitates a power spectrum of two coherent oscillations and is similar to the power spectrum of the data of 2013 November 21--26 (the upper frame of Fig.~\ref{figure8}). The lower frame of Fig.~\ref{figure8} shows the power spectrum of the artificial time series, in which the period of the sine wave varied with ${\rm d}P/{\rm d}T=10^{-8}$. This power spectrum is similar to the power spectrum of a single coherent oscillation and is different from the power spectrum of the data of 2013 November 21--26. Obviously, such period changes are too slow and give no effect within five nearly consecutive nights. Thus, the low degree of coherence of the 67.28-s oscillation is probably caused by phase fluctuations with a standard deviation of about 20\% of the period from night to night.

\begin{figure}[t]
\resizebox{\hsize}{!}{\includegraphics{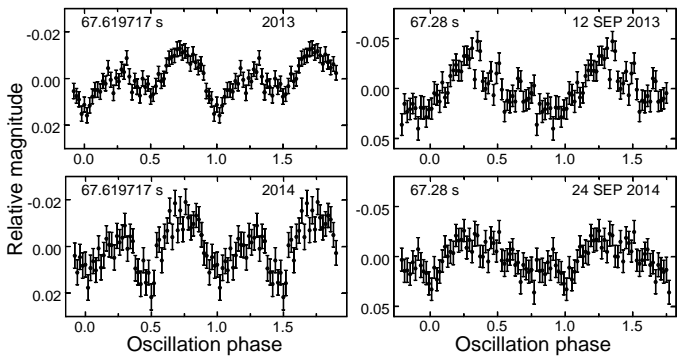}}
\caption{Light curves of \astrobj{V455 And} folded with a period of 67.619717~s (on the left) and with a period of 67.28~s (on the right).}
\label{figure9}
\end{figure}

To find the pulse profile of the coherent oscillation with a period of 67.619717~s, we folded the light curves of \astrobj{V455 And}. The folded light curves were obtained using the data of 2013 and in 2014 separately. As seen in Fig.~\ref{figure9} on the left, both light curves reveal remarkable similarity showing two humps with different heights. This means that the pulse profile is stable. Such a pulse profile is consistent with the average power spectrum which shows the small peak at the 67.61-s period (the insert c in Fig.~\ref{figure2}) and the large peak at its first harmonic (the insert d in Fig.~\ref{figure2}). 

Because the 67.28-s period is of a low degree of coherence, a large portion of data cannot be folded with this period. However, the 67.28-s oscillation has relatively large amplitude and is coherent at least within a night (see Fig.~\ref{figure2}). Therefore, we folded the longest individual light curves separately. Two typical folded light curves are presented in Fig.~\ref{figure9} on the right. As seen, the pulse profile is saw-tooth with a fast increase to maximum and a slow decline to minimum, where its steepness and amplitude are somewhat changeable from night to night. Obviously, an oscillation with such a pulse profile must show only a weak high-frequency harmonic. This is consistent with the average power spectrum (the inserts c and d in Fig.~\ref{figure2}). Thus, the 67.28-s oscillation reveals both changeable phase and changeable pulse profile.

Table~\ref{table2} summarizes information about the periods of the coherent oscillations, which are found using the power spectra. It also includes the approximate peak-to-peak amplitudes which are seen in the folded light curves.

\begin{table}

\caption[ ]{Periods and amplitudes of the detected coherent oscillations.} 

\footnotesize

\label{table2}
\begin{flushleft}
\begin{tabular}{l c c c c }
\noalign{\smallskip}
\hline
\noalign{\smallskip}
\multicolumn{5}{c}{\hspace{1.7cm} 2013 \hspace{2.5cm}   2014} \\
\cline{2-5}
Oscillation    & Period                                      & Ampl.                  & Period                                   & Ampl. \\
                    &                                                  & (mag.)              &                                              & (mag.)   \\                     
\hline
Orbital          & {\scriptsize 81.0862(8) min}   & {\scriptsize 0.160}  & {\scriptsize 81.080(20) min} & {\scriptsize 0.190} \\ 
Superhump & {\scriptsize 80.376(3) min}     & {\scriptsize 0.085}    & {\scriptsize 80.39(7) min}   & {\scriptsize 0.090} \\
Spin             & {\scriptsize 67.619685(40) s}  & {\scriptsize 0.026}   & {\scriptsize 67.6192(13)  s}   & {\scriptsize 0.028}   \\
Harmonic      & {\scriptsize 33.8098585(70) s} &   --                      & {\scriptsize 33.81001(18) s}  &  --  \\

\hline
\end{tabular}
\end{flushleft}
\end{table}

\section{Discussion}

We performed extensive photometric observations of \astrobj{V455~And} with a total duration of 96 h in 2013 and 2014, and clearly found three coherent oscillations with periods of $80.376\pm0.003$~min, $40.5431\pm0.0004$~min and $67.619685\pm0.000040$~s. The shortest period can be the spin period of the white dwarf. The 40.5431-minute period is the first harmonic of the orbital period. Two of these oscillations were observed earlier \citep[e.g.,][]{araujo05, gansicke07}, but the coherent oscillation with a period of 80.376~min was detected by us for the first time. In addition to these coherent oscillations, we found the oscillation with a period of $67.28\pm0.03$~s, which was of a low degree of coherence, and the oscillations with periods in the range 5--6 min, which were incoherent. 

The detection of the oscillation with a period of 80.376~min seems very important because this oscillation can be a negative superhump. We found $P_{\rm orb}= 81.0862\pm0.0008$~min. This $P_{\rm orb}$ conforms to the $P_{\rm orb}$ found by \citeauthor{araujo05} The 80.376-minute period is 0.9\% less than $P_{\rm orb}$ and, hence, this period can be treated as a negative superhump period. In the well-known catalogue of CVs by Ritter and Kolb (\citeyear{ritter03}), \astrobj{V455~And} is classified as a dwarf nova of the WZ~Sge-type. This catalogue contains 83 CVs of the WZ~Sge-type, and none of them is classified as a system showing negative superhumps. Hence, the first detection of negative superhumps in \astrobj{V455~And} is also the first detection of negative superhumps in a WZ~Sge-type dwarf nova. According to the observed magnitude of \astrobj{V455~And} (roughly 16~mag), these superhumps were observed in quiescence.

A negative superhump is usually explained as the beat between the orbital period and the nodal precession of an accretion disc tilted to the orbital plane. To date, no consensus has been found regarding the origin of a disc tilt \citep[e.g.,][and references therein]{montgomery09}. Moreover, in the Ritter-Kolb catalogue we found 43 CVs, which shows negative superhumps. Their types are spread from bright nova-like variables up to dim dwarf novae in quiescence. Thus, our detection of negative superhumps in \astrobj{V455~And} does not contradict existing conceptions.

A positive superhump is usually explained as the beat between the binary motion and the precession of an eccentric accretion disc in the apsidal plane. In contrast to negative superhumps, positive superhumps require high luminosity states and, therefore, are well observed in nova-like variables and dwarf novae superoutbursts and are absent in quiescent dwarf novae \citep{patterson05}. WZ Sge stars in quiescence are the faintest amongst CVs \citep[e.g.,][]{patterson05}. \citet{araujo05}, however, reported the possibility of a positive superhump with a 83.38-minute period in the quiescence of \astrobj{V455 And}. In view of our observations, the existence of these positive superhumps seems doubtful.

We notice that half of the period of 83.38 min claimed by Araujo Betancor et al. is the one-day beat period of the first harmonic of the orbital period of 81.0862 min. We also made numerical experiments simulating the observations by \citeauthor{araujo05} and found that the principal peak in the power spectrum of a coherent oscillation is only 20\% higher than numerous aliases surrounding this peak. This is the effect of extremely large gaps between the observations (see table 1 in \citeauthor{araujo05}). In the presence of an appreciable noise level, the principal peak is inconspicuous, and the fine structure of the window function cannot be traced. In addition, our numerical experiments showed that noise in itself can produce structures resembling the window function. This is also the effect of extremely large gaps between the observations. Thus, both the folded light curve and the power spectrum, which are presented in \citeauthor{araujo05}, give no convincing proofs of the reality of the 83.38-minute oscillation.

The light curves folded with $P_{\rm orb}$ show two humps having nearly equal heights (Fig.~\ref{figure5}). Such profiles of orbital variability were observed in \astrobj{WZ Sge} itself and other WZ Sge stars. Such a profile of orbital variability in \astrobj{V455 And} was observed by \citeauthor{araujo05} They accounted for this profile by two bright spots in the accretion disc. As seen in Fig.~\ref{figure5}, the profile of orbital variability obtained from large portions of data shows only little variations. The individual light curves, however, reveal appreciable differences in different nights. Nonetheless, the adjacent variability cycles with a period of about 80 min seem quite similar to each other (Fig.~\ref{figure1}). This can be explained by the beat between $P_{\rm orb}$ and $P_{\rm sh}$ because the corresponding oscillations have similar amplitudes but different pulse profiles (compare the left and right parts of Fig.~\ref{figure5}). Moreover, the beat period between $P_{\rm orb}$ and $P_{\rm sh}$ is equal to 6.4~d. Hence, the phase difference between these two oscillations remains roughly constant within an individual light curve.
 
Incoherent oscillations with periods of 5--6~min are obvious in our power spectra. These incoherent oscillations were also observed by \citet{araujo05} and \citet{gansicke07}. They account for these oscillations by non-radial pulsations of the white dwarf. \citet{silvestri12} noted that the periods of these oscillations were appreciably shorter (4--5~min) 3 years after the superoutburst which happened in 2007, and the shortening of the periods could be caused by the heating of the white dwarf. We notice that the periods of these oscillations are in the initial range 5--6~min during our observations. The lengthening of the periods can be caused by the cooling of the white dwarf because other outbursts in \astrobj{V455~And} did not happen (see the AAVSO light curve of \astrobj{V455~And}, http://www.aavso.org/data/lcg).

Although not all the periods of \astrobj{V455 And} are fully understood, the presence of the coherent 67.6-s period, which can be the spin period of the white dwarf, is beyond doubt \citep[e.g.,][]{bloemen13}. Therefore, in the IP home page (http://asd.gsfc.nasa.gov/Koji.Mukai/iphome) this system is listed among the confirmed IPs. From our observations, we obtained the precise value of $P_{\rm spin}=67.619685\pm0.000040$~s. The first harmonic of this oscillation allowed us to achieve the more precise value of $P_{\rm spin}=67.619717\pm0.000014$~s.

The oscillation with a period of $67.28\pm0.03$~s has low degree of coherence. Initially, this was not recognized by \citet{araujo05}, and this oscillation was erroneously attributed to the spin period of the white dwarf. \citet{gansicke07} discovered that this signal was incoherent because substantial power remained after pre-whitening. We confirmed that this signal is of a low degree of coherence. This can be caused by fluctuations of the oscillation phase with a standard deviation of about 20\% of the period from night to night.

Although the 67.28-s oscillation is of a low degree of coherence, this oscillation is somehow connected to the spin of the white dwarf. Normally, an IP produces a negative orbital sideband from the reprocessing of X-rays at some part of the system that rotates with $P_{\rm orb}$. The period of this orbital sideband is longer than $P_{\rm spin}$. In addition, the amplitude modulation with $P_{\rm orb}$ can produce a positive orbital sideband with the period shorter than $P_{\rm spin}$. The amplitude of this positive orbital sideband must be noticeably less than the amplitude of the oscillation with $P_{\rm spin}$ \citep{warner86}. Showing the 67.28-s oscillation, \astrobj{V455 And} substantially differs from classical IPs because this oscillation is of a low degree of coherence and, moreover, arises as the beat between $P_{\rm spin}$ and the 3.5-h spectroscopic period, which is in no way related to the orbital period. Both the 67.28-s period and the 3.5-h period are of a low degree of coherence. This strengthens the connection between them.

Obviously, the nature of the 67.28-s oscillation is related to the nature of the 3.5-h spectroscopic period. The 67.28-s oscillation cannot arise from the amplitude modulation with the 3.5-h period because it has much higher amplitude than the amplitude of the oscillation with $P_{\rm spin}$. Then, this oscillation may arise from the reprocessing of X-rays by accretion disc structures which rotate with the 3.5-h period. Obviously, these structures must be in retrograde motion with respect to the spin of the white dwarf because the 67.28-s period is shorter than $P_{\rm spin}$.

\citet{leins92} investigated the free precession of white dwarf stars and computed the increase of the free precession due to elastic properties of the star matter. Using this investigation, \citet{tovmassian07} concluded that a white dwarf with a spin period in the range 50--100~s must have a precession period of a few hours. This allowed them to account for the 3.5-h spectroscopic period observed in \astrobj{V455 And} by the interaction of the magnetic field of the precessing white dwarf and the inner parts of the accretion disc. Unfortunately, \citeauthor{tovmassian07} did not consider the possibility to account for the 67.28-s oscillation according to this conception.

The explanation of the 67.28-s oscillation by the free precession seems possible because the free precession of a white dwarf must be retrograde. Indeed, an oblate body must exhibit the retrograde precession because the angular momentum vector and the instantaneous angular velocity vector projected onto the symmetry axis of the body form an obtuse angle. Hence, when the symmetry axis of the body precesses clockwise, its intrinsic rotation occurs counterclockwise, or vice versa. Thus, the cause of the 67.28-s oscillation can consist in the free precession of the white dwarf.

Two-pole disc-fed accretion is believed to be the normal mode of behaviour in IPs. Depending on the sizes and shapes of the accretion curtains, both single-peaked and double-peaked spin pulse profiles can be produced. In IPs with strong magnetic fields two accreting poles can act in phase so that single-picked, roughly sinusoidal pulse profiles can be produced, whereas in IPs with weak magnetic fields two accreting poles can act in antiphase and produce double-peaked spin pulse profiles \citep{norton99}. It is considered that rapidly spinning IPs with $P_{\rm spin}  < 700$~s have weak magnetic fields and therefore usually produce double-peaked pulse profiles \citep{norton99}. The latter case occurs when the geometry allows the two opposite poles to come into view to the observer \citep[e.g., \astrobj{V405 Aur}:][]{evans04}. According to $P_{\rm spin}$, \astrobj{V455 And} must have a weak magnetic field and must show a double-peaked spin pulse profile. This idea agrees with the double-peaked spin pulse profile in \astrobj{V455 And} (Fig.~\ref{figure9} on the left). In contrast, the 67.28-s oscillation shows a single-picked pulse profile (Fig.~\ref{figure9} on the right).  Obviously, to produce the single-picked pulse profile from a reprocessing, one of the poles of the white dwarf must produce the reprocessing beam invisible to the observer.  

\section{Conclusions}

We performed extensive photometric observations of \astrobj{V455~And} over 19 nights in 2013 and 2014. The total duration of the observations was 96~h.

\begin{enumerate}

\item The analysis of these data allowed us to clearly detect three coherent oscillations with periods of $80.376\pm0.003$~min, $40.5431\pm0.0004$~min and $67.619685\pm0.000040$~s;
\item The 67.619685-s period is the spin period of the white dwarf. Its first harmonic gives the more precise value of $P_{\rm spin}=67.619717\pm0.000014$~s;
\item  The 40.5431-minute period is the first harmonic of the orbital period. This is obvious due to eclipses;
\item The 80.376-minute period is the negative superhump period because this period is 0.9\% less than the orbital period. These negative superhumps were observed in quiescence and were persistent both in 2013 and in 2014;
\item We clearly detected an oscillation with a period of 67.28~s, which was of a low degree of coherence. This oscillation conforms to the beat period between the spin period of the white dwarf and the 3.5-h spectroscopic period, which was discovered earlier and can be accounted for by the free precession of the white dwarf;
\item Because the 67.28-s period is shorter than the spin period of the white dwarf and because the free precession of the white dwarf is retrograde, we account for this period by the free precession of the white dwarf;
\item We observed incoherent oscillations with periods in the range 5--6 min, which were observed earlier and accounted for by non-radial pulsations of the white dwarf. Their periods, which were reduced shortly after the superoutburst, returned into their initial range during our observations;

\end{enumerate}

\begin{flushleft}
{\bf{Acknowlegements}}
\end{flushleft}

Main results were obtained by the use of equipment of the unique scientific unit ''Kourovka Astronomical Observatory''. The research was supported by the State in the name of the Ministry of Education and Science of Russian Federation (the unique project identifier RFMEFI59114X0003). This research has made use of the VizieR catalogue access tool, CDS, Strasbourg, France. This research also made use of the NASA Astrophysics Data System (ADS).

\begin{flushleft}
{\bf{References}}
\end{flushleft}

\bibliographystyle{elsarticle-harv}

\end{document}